\documentclass[11pt]{article}
\usepackage{amsmath}
\usepackage{amssymb}
\usepackage{graphicx}
\usepackage{amssymb}

\newcommand{\CSA}{CS (A)}
\newcommand{\bfAB}{{\bf A \cdot B}}
\newcommand{\pa}{\partial}
\newcommand{\munu}{^{\mu \nu}}
\newcommand{\doefunds}{This work was supported in part by 
funds provided by the U.S. Department of Energy under cooperative research agreement
\#DF-FC02-94ER40818.}
\title{\mbox{Lorentz Violation in a Diffeomorphism-Invariant Theory}\footnote{CPT 07, Bloomington IN, August 2007} }
\author{R. Jackiw\\[1ex]
\small\it Department of Physics\\[1ex]
\small \it Massachusetts Institute of Technology\\[1ex]
\small\it Cambridge, MA 02139\\[1ex]
\small \tt MIT/CTP-3867}
\date{}                                           

\begin{document}

\maketitle

\begin{abstract}
In a diffeomorphism invariant theory, symmetry breaking may be a mask for coordinate choice.
\end{abstract}

\section*{Gauge Theory Preliminary}
My collaborators and I introduced a Lorentz violating modification of the Maxwell theory many years ago \cite{carroll90}, long before Alan Kostelecky constructed a research program on this topic. We were not systematically studying a reasonable class of Lorentz violating interactions, like in Kostelecky's program. Rather we focused on a particular term, which possesses a fine pedigree in physics and mathematics: we added to the Maxwell Lagrange density the quantity 
\begin{equation}
\CSA = \frac{1}{4} \ \varepsilon^{ijk}\, F_{ij}\, A_k = \frac{1}{2} \ \bfAB ,
\label{jackeq1}
\end{equation}
which in mathematics is known as the Chern-Simons term and in physics  is recognized as the Gauss linking number (density). Thus we modified the action to read
\begin{equation}
I = \int d^4 x \ \left[-\frac{1}{4}\, F^{\mu\nu}  F_{\mu\nu} + \frac{m}{2}\ \bfAB\right] ,
\label{jackeq2}
\end{equation}
where $m$ sets the strength of the modification. The dimensional mismatch between the 4-dimensional Maxwell term and the 3-dimensional Chern-Simons term is responsible for the Lorentz (and CTP) violation. An alternative presentation of the action makes use of a time-like embedding vector $v_\mu = (m, {\bf 0})$.
\begin{equation}
I = \int d^4 x \ \left(-\frac{1}{4}\, F^{\mu\nu}  F_{\mu\nu}  + \frac{1}{4}\ v_\mu \, \varepsilon^{\mu\alpha\beta\gamma}\, F_{\alpha \beta}\, A_\gamma\right)
\label{jackeq3}
\end{equation}
Now Lorentz non-invariance is attributed to the presence of the external embedding vector $v_\mu$, which picks out a ``direction" in space-time. Finally the covariant vector present in \eqref{jackeq3}
\begin{eqnarray}
K^\mu &\equiv& \frac{1}{2}\ \varepsilon^{\mu\alpha\beta\gamma}\, F_{\alpha \beta} \, A_\gamma = \varepsilon^{\mu\alpha\beta\gamma}\,\partial_\alpha\, A_ \beta \, A_\gamma\nonumber\\
&&= {^\ast\! F}^{\mu\gamma} A_\gamma\label{jackeq4}\\
\bigg({^\ast\! F^{\mu\nu}} &\equiv& \frac{1}{2}\ \varepsilon^{\mu\nu\alpha\beta}\, F_{\alpha \beta} \bigg)\nonumber
\end{eqnarray}
is recognized as the Chern-Simons current, whose divergence leads to the mathematically/physically significant Chern-Pontryagin density/axial anomaly.
\begin{equation}
\partial_\mu K^\mu = \frac{1}{2}\ {^\ast F^{\mu\nu}  F_{\mu\nu}}
\label{jackeq5}
\end{equation}
Consequently another presentation of the modified action uses an externally prescribed function of time
$\theta = m t$, so that its gradient is our embedding vector $v_\mu = \partial_\mu \theta = (m, {\bf 0})$. After partial integration the action \eqref{jackeq2} or \eqref{jackeq3} reads
\begin{eqnarray}
I &=& \int d^4 x \ \left(-\frac{1}{4}\, F^{\mu\nu}\, F_{\mu\nu} -\frac{1}{4}\, \theta {^\ast F^{\mu\nu}} F_{\mu\nu}\right)\nonumber\\
&=&  \int d^4 x \  \left(-\frac{1}{4}\, F_{\mu\nu} \, M_{\mu\nu\alpha\beta}\, F^{\alpha\beta}\right)\label{jackeq6}\\
M_{\mu\nu\alpha\beta} &\equiv& g_{\mu\alpha} g_{\nu\beta} + \frac{1}{2}\, \varepsilon_{\mu\nu\alpha\beta} \, \theta\nonumber
\end{eqnarray}
and Lorentz non-invariance is attributed to the externally prescribed $\theta$. 

While there is ample formal evidence for Lorentz violation, it is useful to deduce physical consequences of the modified theory and to recognize Lorentz violation in its physical predictions. The equations of the modified theory coincide with the usual Maxwell's equation, except for Amp\`{e}re's law, in which the source electromagnetic current ${\bf J}$ is modified by the addition of the magnetic field: ${\bf J} \to {\bf J} + m {\bf B}$. (Such an alteration is familiar in plasma physics; here it is offered as a change of fundamental physics.) The equations are easily analyzed, and two important physical consequences emerge: gauge invariance is retained, so the ``photon" possesses two polarizations, which propagate in vacuum with velocities that differ from c --- the velocity of light --- and from each other.  The former is a  clear indication of Lorentz non-invariance; the latter signals violation of parity. This causes the vacuum to behave as birefringent  medium and light undergoes a Faraday-like rotation. The modification has a clear observational signature; it has been looked for and none was found. This brings to a very satisfactory conclusion the analysis of a Chern-Simons modification to Maxwell theory.

Before passing to my discussion of the gravity theory modification, let me record the modified Maxwell equations, for later comparison with the gravitational situation
\begin{equation}
J^\nu = \partial_\mu \, F^{\mu\nu} + v_\mu {^\ast\! F^{\mu\nu}}
\label{jackeq7}
\end{equation}
The left side is the conserved current arising from gauge invariant couplings to charged sources, the first term on the right is the variation of the usual Maxwell action; the second comes from varying the Chern-Simons contribution. Taking one further divergence causes each term to vanish separately: $\partial_\nu J^\nu$ because the current is conserved; $\partial_\nu \partial_\mu F^{\mu\nu}$ because $ F^{\mu\nu}$  is anti-symmetric; $v_\mu \partial_\nu {^\ast F^{\mu\nu}}$ because $F^{\mu\nu}$ satisfies a Bianchi identity. Thus no further information is contained in the longitudinal component of the equation of motion.

\section*{Diffeomorphism Invariant Gravity Theory}
It is widely appreciated that a diffeomorphism invariant theory contains structures that parallel analogous quantities in a gauge theory, with the Christoffel connection acting like a (non-Abelian) gauge potential.

In particular, there exists in four dimensions a gravitational Chern-Simon current,
\begin{equation}
K^\mu = \varepsilon^{\mu\alpha\beta\gamma}\ \left(\Gamma^{\ \sigma}_{\alpha\ \tau}\ \partial_{\beta} \, \Gamma^{\ \tau}_{\gamma\ \sigma} + \frac{2}{3}\ \Gamma^{\ \sigma}_{\alpha\ \tau}\ \Gamma^{\ \tau}_{\beta \ \eta} \ \Gamma^{\ \eta}_{\gamma\ \sigma}\right),
\label{jackeq8}
\end{equation}
whose divergence is the gravitational Chern-Pontryagin density.
\begin{eqnarray}
\partial_\mu K^\mu = \frac{1}{2}\ {^\ast R^{\sigma}}_{\, \tau}\, R^{\tau}_{\ \sigma \, \mu\nu} \equiv \frac{1}{2}\ {^\ast RR}\nonumber\\
\left({^\ast R^{\sigma}}_{\, \tau}\, {^{\mu\nu}}  \equiv \frac{1}{2} \varepsilon^{\mu\nu\alpha\beta}\, R^{\sigma}_{\ \tau \alpha \beta}\right)
\label{jackeq9}
\end{eqnarray}
Therefore, analogously to the gauge theoretic case, we propose to modify the Einstein-Hilbert action as \cite{pi2003}
\begin{eqnarray}
I &=& \frac{1}{16\pi G} \ \int d^4 x \ \left(\sqrt{-g}\, R + \frac{1}{4}\ \theta\, {^\ast RR}\right)\nonumber\\
 &=& \frac{1}{16\pi G} \ \int d^4 x \ \left(\sqrt{-q}\, R - \frac{1}{2}\  v_\mu\, K^\mu\right)
 \label{jackeq10}
\end{eqnarray}
where $v_\mu = \partial_\mu \theta$ it is taken to be time-like $v_\mu = (\frac{1}{m}, {\bf 0})$; it is the divergence of $\theta = t/m$. Our anticipation was that here also we shall find physical effects that violate Lorentz invariance. But first let us record the equation of motion for the modified theory
\begin{equation}
-8 \pi \, GT^{\mu\nu} = G^{\mu\nu} + C^{\mu \nu}_{\ \ \, \alpha} \, v^\alpha
\label{jackeq11}
\end{equation}
On the left is the covariantly conserved matter energy momentum tensor, arising from covariantly coupled matter. The first right-hand side term is the conventional Einstein tensor, while the second term is the $g_{\mu\nu}$ variation of the Chern-Simons modification.
Next we take the covariant divergence of \eqref{jackeq11}. Since both $D_\mu T^{\mu\nu}$ and $D_\mu G^{\mu\nu}$ vanish, we are left with
\begin{equation}
D_\mu C^{\mu\nu}_{ \ \ \alpha}  \, v^\alpha = 0.
\label{jackeq12}
\end{equation}
But, unlike the electromagnetic analog, the divergence is non-zero; it is identically given by
\begin{equation}
D_\mu \, C^{\mu\nu}_{\ \ \alpha} \, v^\alpha = \frac{1}{8\sqrt{-g}}\ u^\nu\, {^\ast\! RR}.
\label{jackeq13}
\end{equation}
Hence a consistency condition on our theory is that for $u^\nu \ne 0$
\begin{equation}
{^\ast\! RR} = 0.
\label{jackeq14}
\end{equation}

This consistency condition has important consequences. First of all it supresses the symmetry breaking Chern-Simons term in the action --- even though its variation results in a modified equation of motion. Moreover, the following curious observation must be noted. If $\theta$ is an external variable (as above) then $\int d^4 x \theta \, {^\ast RR}$ apparently violates diffeomorphism invariance and Lorentz invariance. However, let us for a moment consider $\theta$ to be a dynamical quantity, which undergoes variations in the various derivations of the dynamics in the theory. Varying the action with respect to $g_{\mu\nu}$ still produces the equation of motion \eqref{jackeq11}. But with dynamical $\theta$, we must also vary that quantity, and we obtain the equation \eqref{jackeq14}. But that equation also emerged with non-dynamical $\theta$, as a subsidiary condition on \eqref{jackeq11}. So it appears that the same equations hold, whether $\theta$ is dynamical or external. It is our purpose to clarify this further.

In order to illuminate the issue of Lorentz violation, we should look to physical consequences, rather than to the confusing formal properties of the theory. Although the non-linear equations are much too complicated to be solved exactly, much has been done by many, at least on the approximate level.
\begin{itemize}
\item[(a)] The classical, solar system test of general relativity survive unchanged \cite{pi2003}. The reason is that they are based on a static, radially symmetric geometry for which the Chern-Simons term vanishes. But Smith {\it et. al.} have found effects of the modification on bodies orbiting the earth \cite{smith:etal}. 

\item[(b)] Since ${^\ast RR}$ must vanish, the Kerr solution, for which ${^\ast\! RR \ne 0}$, must be deformed to accomodate rotating black holes. Some investigation of this is due to Konna {\it et al} \cite{konno:etal}.

\item[(c)] Both polarizations of gravitational waves propagate with velocity $c$, but with different intensity \cite{pi2003}. This puts into evidence parity violation, but not Lorentz violation. Parity violating effects are also established in the post Newtonian expansion by Alexander and Yunes \cite{alex:yun99}, and  in cosmological solutions by Lue {\it et al}. as well as by Alexander \cite{lue:1983}.

\item[(d)] A useful summary of work, also performed in China, is by Ni \cite{ni2005}. 
\end{itemize}

All this gives ample evidence for physical consequences of the Chern-Simons modification, but no persuasive consequence of Lorentz  violation is identified. To illuminate this further, two young researchers at MIT performed a formal analysis of the theory from which a definite conclusion can be drawn.

Let us recall the formal approach to Lorentz and Poincar\'{e} invariance in a Lagrangian ($\mathcal{L}$) field theory for a field $\varphi$. ($\varphi$ need not be a scalar field, but its vectorial/fermionic indices are suppressed; in fact in the present application $\varphi$ stands for the metric tensor $g_{\alpha \beta}$.)

An infinitesimal Poincar\'{e} transformation of the coordinates $ x^\alpha \to x^{\prime \alpha} = x^\alpha - a^\alpha - \omega^\alpha\  _\beta\ x^\beta$ is accompanied by the field transformation
\begin{equation}
\varphi \to \varphi^\prime ;\ \  \varphi^\prime (x^\prime) = \varphi (x) + \frac{1}{2} \ \omega^\alpha \, _\beta\ S^\beta\, _\alpha \ \varphi (x)
\label{jackeq15}
\end{equation}
where $\omega^0 \, _0 =0, \, \omega^0 \, _i  =\omega^i \, _0 =  0, \omega^i \,  _j = -\omega^j \,  _i$ while $S^\alpha \, _\beta$ represents the Lorentz group on the field.
\begin{equation}
S^0 \, _0 = 0, \ \ S^0 \, _i = S^i \, _0 = 0, \ \ S^i \, _j = - S^j \, _i
\label{jackeq16}
\end{equation}

Invariance of the action against translations $(a^\alpha \ne 0, \omega^\alpha \, _\beta = 0$) leads by Noether's theorem to a two-index conserved ``tensor" $\theta^\mu \, _\alpha, \partial_\mu\, \theta^\mu \, _\alpha = 0$. When the action is also Lorentz invariant ($\omega^\alpha \,\! _\beta \ne 0$) one can show that
\begin{equation}
\theta^\mu \,\! _\alpha = \theta^\mu \,\! _{B \alpha} + \partial_\nu\, X^{[\nu, \mu]} \, _\alpha
\label{jackeq17}
\end{equation}
where $X^{[\nu, \mu]} \, _\alpha$ is antisymmetric in $[\nu, \mu]$ and $\theta^{\mu\nu}_B \equiv \theta^\mu _{B \alpha} \, \eta^{\alpha\nu}$ is symmetric in ($\mu, \nu$). [$\eta^{\alpha \nu}$ is the flat metric tensor.] Due to the antisymmetry of $X^{[\nu, \mu]} \, _\alpha$, conservation of $\theta^\mu \,\! _\alpha$ implies conservation of $\theta^{\mu\nu}_B$. Thus presence of Poincar\'{e} invariance in a field theoretic action ensures the existence of two-index, symmetric and conserved ``tensor" --- the Belinfante energy-momentum ``tensor." Also vice-versa: the existence of a symmetric, conserved two-index object guarantees invariance agaisnt Poincar\'{e} transformations. (The quotation marks on ``tensor" remind us that quantity is not a generally covariant tensor, but only Poincar\'{e} convarient. Henceforth the distinction will not be made.)

When this information is brought to gravity theory, various complications arise. First, the Lagrangian involves second derivatives of the field, but Noether's theorem is usually presented for first-derivative Lagrangians. In fact this is not a problem for the Einstein-Hilbert action: because of the identity
\begin{eqnarray}
\int d^4 x \, \sqrt{-g}\ R = \int d^4 x \ \sqrt{-g}\ g^{\mu\nu}\ \left(\Gamma^\alpha_{\mu\beta}\, \Gamma^\beta_{\nu\alpha} - \Gamma^\alpha_{\mu\nu}\, \Gamma^\beta_{\alpha \beta}\right)\nonumber\\
+ \ \text{surface term}\hspace{1in}
\label{jackeq18}
\end{eqnarray}
we may use a first-deviative action. But higher derivatives are irremovably present in our modification ${^\ast\! RR}$.

To overcome this problem, Noether's theorem can be extended. One finds for translation invariance
\begin{equation}
\theta^\mu \, _\alpha = \Pi^\mu \partial_\alpha \, \varphi + \Pi^{\mu\nu}\, \partial_\nu\, \partial_\alpha \, \varphi - \partial_\nu \, \Pi^{\mu\nu}\, \partial_\alpha\, \varphi - \delta^\mu_\alpha \mathcal{L}
\label{jackeq19}
\end{equation}
and the equation of motion, which shows that the above is conserved, reads
\begin{eqnarray}
\partial_\mu \Pi^\mu = \frac{\partial \mathcal{L}}{\partial\varphi} + \pa_\mu \pa_\nu \Pi\munu ,\label{jackeq20}\\
\text{where}\ \Pi^\mu \equiv \frac{\pa\mathcal{L}}{\pa\pa_\mu \varphi}\ , \ \ \Pi\munu=\frac{\pa\mathcal{L}}{\pa \pa_\mu \pa_\nu \varphi}.
\label{jackeq21}
\end{eqnarray}

When this procedure was applied to the Einstein-Hilbert action by Papapetrou and later by Bak {\it et al.} \cite{pap1948} the result was 
\begin{equation}
^{EH}\!\theta\munu_B = \frac{1}{8\pi G}\ \pa_\alpha \pa_\beta\ \left[\sqrt{-g}\, \eta^{[\mu} g^{\alpha] \beta} + \eta^{\beta [\alpha} g^{\mu] \nu}\right]
\label{jackeq22}
\end{equation}
Note that in harmonic gauge, $\pa_\alpha (\sqrt{-g}\ g^{\alpha\beta}) = 0$, the above reduces to the simple expression
\[
^{EH}\!\theta\munu_B = \frac{1}{16\pi G}\ \Box\, g\munu
\]

This program can now be repeated for Chern-Simons modified gravity. The tensor arising from translation invariance reads
\begin{equation}
^{CS} \theta^\mu \, _\alpha = \theta^\mu \, _\alpha + \frac{1}{2}\ v_\alpha \, K^\mu
\label{jackeq23}
\end{equation}
$\theta^\mu \, _\alpha$ is constructed as in \eqref{jackeq19}-\eqref{jackeq21}, except that the field derivatives in\eqref{jackeq20}, \eqref{jackeq21} are taken from the complete modified Lagrange density. The analog to \eqref{jackeq17} applies to the above constructed $\theta^\mu \, _\alpha$.
\begin{equation}
\theta^\mu \, _\alpha = \theta^\mu \, _{B\alpha} + \pa_\nu X^{[\nu, \mu]}\, _\alpha + A^\mu \, _\alpha
\label{jackeq24}
\end{equation}
The first two terms are as in \eqref{jackeq17}; the last term depends on $^\ast\! RR$, and vanishes on shell. Therefore, on shell, \eqref{jackeq23} may be presented as
\begin{equation}
^{CS} \theta^\mu \, _\alpha = \theta^\mu \, _{B \alpha} + \pa_\nu\, X^{[\nu, \mu]}\, _\alpha + \frac{1}{2}\ v_\alpha  \, K^\mu
\label{jackeq25}
\end{equation}
Conseervation of $^{CS} \theta^\mu \, _\alpha$ implies that
\begin{equation}
\pa_\mu \theta^\mu \, _{B \alpha} = - \frac{1}{2}\ v_\alpha \pa_\mu K^\mu = - \frac{1}{2}\ v_\alpha {^\ast\! RR} = 0
\label{jackeq26}
\end{equation}
Thus the Chern-Simons modified theory admits a symmetric, conserved two-index tensor, leading to the conclusion that Poincar\'{e} invariance holds with the Chern-Simons modification in place.

\section*{Reprise}
Can we understand {\it a priori} why no Lorentz breaking effects are visible in the Chern-Simons modified model? Here is a possible argument. We have previously remarked that the equations of motion are the same, regardless whether $\theta$ is a dynamically varying field or the externally prescribed field,  $\theta = \frac{1}{m t}$. Also we saw that the equation ${^\ast\! RR} =0$, which is a consistency condition on the equation of motion with $\theta$ external, extinguishes the symmetry breaking term in the action.

Let us now view the fully dynamical theory, with $\theta$ taken to be a dynamical variable. Complete diffeomorphism invariance holds, with $\theta$ transforming as a scalar. But now we can use the diffeomorphism transformation to fix $\theta$ at $\frac{1}{mt}$. In other words, what appears to be a symmetry breaking external field, is in fact merely a choice of coordinates (choice of gauge) in a coordinate invariant (gauge invariant) model.

So I leave you with the following caution. Apparent symmetry breaking may in fact be just a choice of gauge!

\doefunds

\end{document}